\documentclass[aip,reprint,amsmath,asmsymb,groupedaddress]{revtex4-1}

\usepackage{graphicx}
\usepackage{bm}
\usepackage[utf8]{inputenc}
\usepackage{xcolor}

\newcommand*{\xvec}{\mathbf{x}}
\newcommand*{\Dvec}{\mathbf{D}}
\newcommand*{\ED}{\langle \Dvec \rangle}
\newcommand*{\Kmat}{\mathbf{K}}

\newcommand*{\cvec}{\mathbf{c}}
\newcommand*{\Cmat}{\mathbf{\Sigma}}
\newcommand*{\quadratic}[2]{{#2}^t\cdot{#1}\cdot{#2}}

\begin{document}

\title{Molecular simulations minimally restrained by experimental data}

\author{Huafeng Xu}
\email[Correspondence to ]{huafeng.xu@silicontx.com} 
\affiliation{Silicon Therapeutics LLC, Boston, USA}

\begin{abstract}

  One popular approach to incorporating experimental data into molecular
  simulations is to restrain the ensemble average of observables to
  their experimental values.  Here I derive equations for the
  equilibrium distributions generated by restrained ensemble
  simulations and the corresponding expected values of observables.
  My results suggest a method to restrain simulations so that they
  generate distributions that are minimally perturbed from the
  unbiased distributions while reproducing the experimental values of
  the observables within their measurement uncertainties.

\end{abstract}

\pacs{}

\maketitle

\section{Introduction}

Molecular simulations are routinely used to generate models of
molecular structures and their equilibrium
distributions~\cite{dror2012}.  They provide structural and dynamic
information at atomistic resolution that are critical to the
interpretation of experimental data~\cite{bottaro2018}.  Inaccuracy in
the empirical force fields used to compute the potential energy of the
biomolecules in the simulations, however, can cause the simulations to
generate inaccurate equilibrium distributions of molecular structures,
manifested by the disparity between the experimentally measured values
of observables and the predictions by the
simulations~\cite{gunsteren2008molecular}.  Biasing the simulations
toward reproducing the experimental values of the observables is a
promising approach to improving the accuracy of the
simulations~\cite{Boomsma:2014b,Perez:2016}.  How to introduce minimal
bias, however, remains an open question, and no such method exists
that takes into account the measurement uncertainties associated with
the experimental data.

Because many experimental techniques, such as nuclear magnetic
resonance (NMR) and small-angle X-ray scattering (SAXS), measure the
ensemble average of observables, a natural and popular method to
incorporate such experimental data is to simultaneously simulate many
replicas of the molecular system while restraining the average of the
instantaneous values of any observable of interest to be close to its
experimental
value~\cite{Lindorff-Larsen:20056f8,Vendruscolo:20076f8,Roux:2013a,hummer2015}.
I will refer to such replica-based restrained simulations simply as
restrained ensemble simulations.  The equilibrium distribution
generated by restrained ensemble simulations depends on the number of
replicas and the strength of the restraints, the latter of which
should reflect the measurement uncertainties of the
observables~\cite{hummer2015}.  There lacks, however, 
a clear theoretical underpinning for the choice of the number of 
replicas~\cite{olsson2015}.

The restrained simulations should be no more biased by the restraints
than is necessary, which is a principle that motivated the alternative
maximum entropy method, which seeks a distribution that reproduces the
experimental values of the observables and is minimally perturbed from
the unbiased distribution
~\cite{Boomsma:2014,Rozycki:20116f8,Groth:19996f8,Massad:20076f8}. It
has been established that in some limit of infinite number of replicas
and infinitely strong restraints, restrained ensemble simulations
generate a statistically equivalent distribution as the maximum
entropy
method~\cite{pitera2012,Roux:20136f8,Cavalli:20136f8,olsson2015}.
In practice, however, the restraining strength is finite, in which case
the two methods are not equivalent. In contrast to restrained
ensemble simulations, the maximum entropy method constrains every
observable to its exact experimental value, thus it does not take into
account the measurement uncertainties, which are ubiquitous in all
experiments.

Here I present theoretical results that enable the determination of
the largest number of replicas to use in restrained ensemble
simulations with fixed strength of restraints so that they generate a
distribution minimally perturbed from the unbiased distribution while
reproducing the experimental values of the observables within their
measurement uncertainties.  As I will demonstrate below, this is
equivalent to the determination of the weakest strength of restraints
to use at fixed number of replicas, provided that the latter is
sufficiently large. The key result is an equation
(Eq.~\ref{eqn:D-expected}) that predicts the expected value of
any--restrained and unrestrained--observable in restrained ensemble
simulations for different numbers of replicas, thus permitting the
selection of the appropriate number without computationally expensive
trial and error. In addition, I derive an equation
(Eq.~\ref{eqn:marginal-distribution}) for the distribution generated
by restrained ensemble simulations with a finite number of replicas
and a finite restraining strength, which can be used to perform
reweighted simulations that take into account of the experimental data
with their attendant measurement uncertainties.  The validity of these
theoretical results are demonstrated numerically by Monte Carlo
simulations. I illustrate the application of these results using the
molecular dynamics (MD) simulations of the polyalanine peptide Ala$_5$
restrained by experimental nuclear magnetic resonance (NMR) data.

\section{Theory}

Consider an ensemble consisting of $N$ replicas of the molecular
system, each with the atomic coordinates $\xvec_{i=1,2,\dots,N}$.
Experiments measure the ensemble average of the individual
contributions of the observables: $\bar{\Dvec} =
N^{-1}\sum_i\Dvec_{i}$, where $\Dvec_i \equiv \Dvec(\xvec_i) =
(D_1(\xvec_i), D_2(\xvec_i), \dots, D_M(\xvec_i)^t$ are the
instantaneous values of the vector of $M$ observables for the $i$'th
replica.  In an unbiased simulation, $\bar{\Dvec}$ may deviate from
experimental values $\Dvec_e$.  The posterior distribution given the
experimental data~\cite{Rieping2005,hummer2015} is
\begin{eqnarray}
p_N(\{\xvec_{i=1,2,\dots,N}\}| \Dvec_e) &=& p(\{\xvec_i\}|\Dvec_e,E) 
\nonumber \\
&\propto&
  p(\Dvec_e|\{\xvec_i\})p(\{\xvec_i\}|E)
\nonumber \\
&=& p(\Dvec_e|\{\xvec_i\}) \prod_i p_0( \xvec_i)
\label{eqn:Bayesian-posterior}
\end{eqnarray}
where $p_0(\xvec_i) = Z_0^{-1} \exp( -\beta E(\xvec_i))$ is the
Boltzmann distribution of the molecular configurations under the
potential energy function $E$.  Assuming a normal distribution in each
experimental value, $p(\Dvec_e|\{\xvec_i\}) = (\prod_\alpha 2\pi
\delta_\alpha^2)^{-M/2}
\exp(-\frac{1}{2}\quadratic{\Kmat}{(\bar{\Dvec} - \Dvec_e)})$, where
$\Kmat = \text{diag}(\delta_1^{-2}, \delta_2^{-2}, \dots,
\delta_M^{-2})$ is the inverse of the statistical variances of the
experimental measurements, $p_N(\{\xvec_i\}|\Dvec_e)$ becomes
\begin{equation}
p_N(\{\xvec_i\}; \Kmat, \Dvec_e) \propto 
e^{ -\sum_i \beta E(\xvec_i) - \frac{1}{2}\quadratic{\Kmat}{(\bar{\Dvec} - \Dvec_e)} },
\label{eqn:restrained-ensemble-distribution}
\end{equation}
which is the distribution in the restrained ensemble simulations.  The
strength of the restraints, $\Kmat$, is thus related to the
measurement uncertainties associated with the experimental data.
In the restrained ensemble simulations, $N$ replicas of the
molecular system are simulated with the same energy function and at
the same temperature, and they interact with one another only
through the quadratic term in the exponent in
Eq.~\ref{eqn:restrained-ensemble-distribution}.

A similar equation was derived by Hummer and K\"{o}finger
(Eq. 26 of Ref.~\onlinecite{hummer2015}), with the difference that
they scaled the restraining term by $N$ and introduced a parameter
($\theta$) expressing the level of confidence in the unbiased
reference distribution.  My results below show that these two equations 
are equivalent at large $N$.  

For a fixed $\Kmat$, the marginal distribution for a single replica,
$p_N(\xvec_1; \Kmat, \Dvec) = \int \prod_{i=2}^N p_N(\{ \xvec_i \};
\Kmat, \Dvec_e)$, tends to the unbiased distribution $p_0(\xvec_1)$
with increasing $N$.  Larger $N$ thus corresponds to higher
confidence in the unbiased distribution.  A reasonable choice of $N$ is
to take the largest value such that the expected values of the
observables in the restrained ensemble simulations remain within
the measurement uncertainties of the experimental values: this generates a
distribution minimally perturbed from the unbiased distribution while
still reproducing the experimental data.

I will analyze the behavior of restrained ensemble simulations at
sufficiently large numbers of replicas, {\it i.e.} $N\gg 1$. Let
\begin{equation}
p_\lambda(\xvec) = Z_\lambda^{-1} \exp(-\beta E(\xvec) +
\lambda^t\cdot \Dvec(\xvec))
\label{eqn:biased-distribution}
\end{equation}
be a biased distribution for an arbitrary vector $\lambda$, which
becomes the distribution generated by the maximum entropy method if
$\lambda$ is the corresponding Lagrangian
multiplier~\cite{Roux:20136f8}.  Dropping the subscript from the
constant vector $\Dvec_e$, the partition function for the restrained
ensemble of $N$ replicas is
\begin{eqnarray}
&&Z_N(\Kmat, \Dvec) 
\nonumber \\
&=& \int \prod_i d\xvec_i e^{ 
  - \sum_i \beta E(\xvec_i)
 - \frac{1}{2}\quadratic{\Kmat}{( N^{-1}\sum_i \Dvec_i - \Dvec)}}
\nonumber \\
&=& Z_\lambda^N \langle 
 e^{ -\frac{1}{2} \quadratic{\Kmat}{(N^{-1}\sum_i \Dvec_i - \Dvec)} - \lambda^t\cdot \sum_i \Dvec_i} \rangle_{N, \lambda}
\nonumber \\
&=& Z_\lambda^N \int d\bar{\Dvec} ( \langle \delta( N^{-1}\sum_i \Dvec_i - \bar{\Dvec}) \rangle_{N,\lambda} \cdot
\nonumber \\
 && e^{-\frac{1}{2}\quadratic{\Kmat}{(\bar{\Dvec} - \Dvec)} - N\lambda^t\cdot \bar{\Dvec}} )
\nonumber \\
&=& Z_\lambda^N \int d\bar{\Dvec} 
e^{-\frac{1}{2}\quadratic{\Kmat}{(\bar{\Dvec} - \Dvec)} - N\lambda^t\cdot \bar{\Dvec}} 
\rho_{N,\lambda}(\bar{\Dvec})
\label{eqn:ZN}
\end{eqnarray}
where $\langle \cdots \rangle_{N,\lambda}$ signifies the ensemble
average according to the probability distribution of $N$ independent
replicas ({\it i.e.,} $\Kmat = \mathbf{0}$), each sampled according to
the biased distribution $p_\lambda(\xvec)$, and
$\rho_{N,\lambda}(\bar{\Dvec})$ is the corresponding probability
distribution of the random variable $\bar{\Dvec} = N^{-1}\sum_i
\Dvec_i$.  In this ensemble of $N$ independent replicas, $\{\Dvec_i\}$
are independent and identically distributed (i.i.d.) random variables.
Acccording to the central limit theorem, the random variable
$\bar{\Dvec} = N^{-1}\sum_i \Dvec_i$, for sufficiently large $N$,
tends to the normal distribution $\rho_{N,\lambda}(\bar{\Dvec})
\approx (2^M\pi^M|\Cmat_\lambda|/N)^{-1/2} e^{-\frac{1}{2}\quadratic{
    N\Cmat^{-1}}{(\bar{\Dvec} - \ED_\lambda)}}$, where $\ED_\lambda =
\int d\xvec\, \Dvec(\xvec) p_\lambda(\xvec)$ are the expected values of
the observables $\Dvec(\xvec)$ in the biased distribution (I will use
$\langle \cdots \rangle_\lambda$ to denote the
expected values in the biased distribution with parameter $\lambda$),
$\Cmat_\lambda = \langle (\Dvec(\xvec) - \ED)\cdot(\Dvec(\xvec) -
\ED)^t \rangle_\lambda$ is the corresponding covariance matrix, and
$|\Cmat_\lambda|$ is the determinant of matrix $\Cmat_\lambda$.  If
$\Dvec_\lambda \equiv \Dvec - N\Kmat^{-1}\cdot\lambda$ and $\ED_\lambda$ are close, in the sense that 
\begin{equation}
\max_{\substack{\bar{\Dvec} \ s.t. \\  e^{-\quadratic{\Kmat}{(\bar{\Dvec} - \Dvec_\lambda)}} > \epsilon}} \quadratic{N\Cmat^{-1}}{(\bar{\Dvec} - \ED_\lambda)} < \alpha
\label{eqn:average-D-overlap}
\end{equation}
for a sufficiently small $\epsilon > 0$ and some value of $\alpha\geq 3$,
such that the normal distribution is a good approximation to
$\rho_{N,\lambda}(\bar{\Dvec})$ for all $\bar{\Dvec}$ around
$\Dvec_\lambda$, $\bar{\Dvec}$ can be integrated out, and the
partition function in Eq.~\ref{eqn:ZN} becomes
\begin{eqnarray}
Z_N(\Kmat, \Dvec) &\approx& \frac{Z_\lambda^N e^{\frac{1}{2}N^2\quadratic{({\Kmat + N\Cmat_\lambda^{-1}})^{-1}}{\lambda}}}{\sqrt{|I + N^{-1}\Kmat\cdot\Cmat_\lambda|}}
\nonumber \\
&\cdot& e^{-\frac{1}{2}\quadratic{(\Kmat^{-1} + N^{-1}\Cmat)^{-1}}{(\Dvec - \ED_\lambda)}}
\nonumber \\
&\cdot& e^{-N\lambda^t\cdot(\Kmat+N\Cmat_\lambda^{-1})^{-1}\cdot(\Kmat\cdot\Dvec + N\Cmat_\lambda^{-1}\cdot\ED_\lambda)}
\label{eqn:ZNKD}
\end{eqnarray}
where $I$ is the identity matrix.  Eq.~\ref{eqn:ZNKD} forms the basis for deriving the key results of this work.

My first key result is that the expected value of any observable,
restrained or unrestrained, in the restrained ensemble simulations can
be predicted from the biased simulation of $p_\lambda(\xvec)$.  Let
$\Dvec^\ast = (D_1, D_2, \dots, D_M, D_{M+1}, \dots, D_{M+M'})^t$ be
the vector of restrained and unrestrained observables, where $\Dvec =
(D_1, D_2, \dots, D_M)$ are the $M$ restrained observables and
$\Dvec_0 = (D_{M+1}, \dots, D_{M+M'})^t$ are the $M'$ unrestrained
observables. Introducing
\begin{equation}
\Kmat^\ast = \left(\begin{array}{ll}
\Kmat & \mathbf{0}_{M\times M'} \\
\mathbf{0}_{M'\times M} & \mathbf{0}_{M'\times M'}
\end{array}\right),
\end{equation}
$\Dvec_e^\ast = (\Dvec_e^t\ \, \mathbf{0}_{1\times M'})^t$, and
$\lambda^\ast = (\lambda^t\ \, \mathbf{0}_{1\times M'})^t$, the expected
values of the observables $\Dvec^\ast(\xvec)$ in the restrained
ensemble simulations are given by
\begin{eqnarray}
&&\langle \Dvec^\ast \rangle_{N,\Kmat,\Dvec_e} 
\nonumber \\ 
&=&
\int \prod_i d\xvec_i p_N(\{\xvec_i\};\Kmat, \Dvec_e) N^{-1}\sum_i \Dvec^\ast(\xvec_i)
\nonumber \\
&=& \frac{\partial}{\partial\mathbf{\gamma}} \ln\left(
  \int \prod_i d\xvec_i ( e^{-\beta\sum_i E(\xvec_i)} \right.
\nonumber \\
&&\cdot \left. \left. e^{-\frac{1}{2}\quadratic{\Kmat}{(N^{-1}\sum_i \Dvec_i - \Dvec_e)}} \right. \right.
\nonumber \\
&&\cdot \left. \left. e^{\mathbf{\gamma}^t\cdot N^{-1} \sum_i \Dvec^\ast_i})\right) \right|_{\mathbf{\gamma} = 0}
\nonumber \\
&=& \left. \frac{\partial}{\partial\mathbf{\gamma}} \ln \left(Z_N(\Kmat, \Dvec_e) e^{\frac{1}{2}\quadratic{(\Kmat^\ast + N\Cmat_\lambda^{\ast -1})^{-1}}{\gamma}} \right. \right.
\nonumber \\
&\cdot& \left. \left. e^{\gamma^t\cdot(\Kmat^\ast + N\Cmat^{\ast -1})^{-1}\cdot(\Kmat^\ast\cdot
\Dvec_e^\ast
 + N(\Cmat_\lambda^{\ast -1}\cdot\langle \Dvec^\ast \rangle_\lambda - \lambda^\ast) }\right)\right|_{\gamma=0}
\nonumber \\
&=& 
\Dvec_e^\ast + (I + N^{-1}\Cmat^\ast_\lambda\Kmat^\ast)^{-1}\cdot( \langle \Dvec^\ast \rangle_\lambda - \Dvec_e^\ast
- \Cmat^\ast_\lambda\cdot \lambda^\ast)
\nonumber \\
\label{eqn:D-expected}
\end{eqnarray}
where $\Cmat^\ast_\lambda = \langle (\Dvec^\ast - \langle \Dvec^\ast
\rangle)\cdot(\Dvec^\ast - \langle \Dvec^\ast \rangle)^t
\rangle_\lambda$ is the covariance between all the observables in the
biased simulation.
 
Writing $\Cmat^\ast$ in the block form
\begin{equation}
\Cmat_\lambda^\ast = \left(
\begin{array}{cc}
\Cmat_\lambda & \cvec_{0,\lambda}^t \\
\cvec_{0,\lambda} & \Cmat_{00,\lambda}
\end{array}\right)
\end{equation}
where $\Cmat_{00,\lambda}$ is the $M'\times M'$ covariance matrix for
the unrestrained observables $\Dvec_0$ and $\cvec_{0,\lambda}$ is the
$M'\times M$ covariance matrix between the unrestrained and the restrained
observables, the expected values of the restrained observables and the
unrestrained observables are, respectively,
\begin{eqnarray}
&&\langle \Dvec(\xvec_1) \rangle_{N,\Kmat,\Dvec_e} 
\nonumber \\
&=& \Dvec_e 
+ (I + N^{-1}\Cmat_\lambda\Kmat)^{-1}\cdot(\ED_\lambda - \Dvec_e - \Cmat_\lambda \cdot \lambda)
\nonumber \\
\label{eqn:restrained-D-expected}
\end{eqnarray}
and
\begin{eqnarray}
&& \langle \Dvec_0(\xvec_1) \rangle_{N,\Kmat,\Dvec_e} 
\nonumber \\
&=& \langle \Dvec_0 \rangle_\lambda  
+ \cvec_{0,\lambda}\cdot\Cmat_\lambda^{-1}(I + N\Kmat^{-1}\cdot\Cmat_\lambda^{-1})^{-1}\cdot( \Dvec_e - \ED_\lambda)
\nonumber \\
&+& \cvec_{0,\lambda}\cdot\Cmat_\lambda^{-1}\cdot((I+N\Kmat^{-1}\cdot\Cmat^{-1})^{-1}-I)\cdot\lambda
\label{eqn:unrestrained-D-expected}
\end{eqnarray}

Eq.~\ref{eqn:restrained-D-expected} enables the determination of the
largest $N$ such that the expected values of the restrained
observables remain within the measurement uncertainties of their
experimental values.

My second key result is that the marginal distribution for a single
replica generated by the restrained ensemble simulations is
\begin{eqnarray}
&&p_N(\xvec_1; \Kmat, \Dvec) 
\nonumber \\
&=& Z_N^{-1}(\Kmat, \Dvec)e^{-\beta E(\xvec_1)} 
  \int \prod_{i=2}^{N} d\xvec_i \left( e^{-\beta\sum_{i=2}^{N} E(\xvec_i)} \right.
\nonumber \\
&&\cdot
 \left.  e^{-\frac{1}{2}\quadratic{\Kmat}{\left(N^{-1}\sum_{i=1}^{N} \Dvec_i - \Dvec\right)}}
 \right)
\nonumber \\
&=& Z_N^{-1}(\Kmat,\Dvec) Z_{N-1}(\frac{(N-1)^2}{N^2}\Kmat, \frac{N \Dvec - \Dvec_1}{N-1}) e^{-\beta E(\xvec_1)}
\nonumber \\
&\propto& p_0(\xvec_1) 
e^{(\lambda + \Cmat_\lambda^{-1}\cdot(\Dvec - \ED_\lambda))^t \cdot (I + N\Kmat^{-1}\cdot\Cmat_\lambda^{-1})^{-1}\cdot( \Dvec(\xvec_1) - \Dvec)}
\nonumber \\
&\cdot& e^{-\frac{1}{2N}\quadratic{(N\Kmat^{-1} + \Cmat_\lambda)^{-1}}{(\Dvec(\xvec_1) - \Dvec)}},
\label{eqn:marginal-distribution}
\end{eqnarray}
where I used Eq.~\ref{eqn:ZNKD} for both $Z_N(\Kmat,\Dvec)$ and
$Z_{N-1}(\frac{(N-1)^2}{N^2}\Kmat, \frac{N \Dvec - \Dvec_1}{N-1})$,
and $N\approx N-1$ for large $N$.

For $N\gg 1$, $\frac{1}{2N}\quadratic{(N\Kmat^{-1} +
  \Cmat_\lambda)^{-1}}{(\Dvec_1 - \Dvec)} \approx 0$, and
Eq.~\ref{eqn:marginal-distribution} reduces to
\begin{eqnarray}
&& p_N(\xvec; \Kmat, \Dvec_e) 
\nonumber \\
&\propto& p_0(\xvec) 
  e^{(\lambda + \Cmat_\lambda^{-1}\cdot(\Dvec_e - \ED_\lambda))^t \cdot
  (I + N\Kmat^{-1}\cdot\Cmat_\lambda^{-1})^{-1}\cdot(\Dvec(\xvec) - \Dvec_e)},
\nonumber \\
\label{eqn:marginal-func-of-NiK}
\end{eqnarray}
which depends on $N$ through the product $N\Kmat^{-1}$.  This suggests
that a restrained ensemble simulation with $N$ replicas and the
strength of restraints $\Kmat$ is equivalent to another one with $N'$
replicas and the strength of restraints $\Kmat' = (N'/N) \Kmat$.
This equivalency, to my knowledge, has not been
previously demonstrated except for the harmonic potentials~\cite{Roux:20136f8}, and it
mathematically justifies the scaling by $N$ of the restraining term
proposed by Hummer and K\"{o}finger~\cite{hummer2015}. 

Eq.~\ref{eqn:marginal-distribution} suggests that the restrained ensemble
simulations generate the same distribution as the maximum entropy
method if the number of replicas satisfies the following inequality
\begin{equation}
1 \ll N \ll \min( \text{eigenvalues of }\Kmat\cdot\Cmat_\lambda).
\label{eqn:good-N-range}
\end{equation} 
This condition is the generalization of the condition $N\ll K$ derived
for the special case of harmonic potentials~\cite{Roux:20136f8}.
Under this condition, $(I + N\Kmat^{-1}\Cmat_\lambda^{-1})^{-1}
\approx I$, the
corresponding single-replica distribution becomes
\begin{equation}
p_N(\xvec_1;\Kmat,\Dvec_e) \propto
p_0(\xvec_1) e^{(\lambda + \Cmat_\lambda^{-1}\cdot(\Dvec_e - \ED_\lambda))^t\cdot (\Dvec(\xvec_1) - \Dvec_e)},
\label{eqn:ME-equivalency}
\end{equation}
which is the same as $p_\lambda(\xvec_1)$ if $\lambda$ takes the value
of the Lagrangian multiplier $\lambda_{\text{ME}}$ in the maximum
entropy method such that $\ED_{\lambda_{\text{ME}}} = \Dvec_e$.  Under
this condition, Eq.~\ref{eqn:restrained-D-expected} becomes $\langle
\Dvec(\xvec_1) \rangle_{N,\Kmat,\Dvec_e} \approx \Dvec_e$.

The Kullback-Leibler divergence from the distribution in the
restrained ensemble simulations (Eq.~\ref{eqn:marginal-distribution})
to that of $p_{\lambda'}(\xvec)$ is
\begin{eqnarray}
\Delta_{\text{KL}}[p_{\lambda'}||p_N] &=&
  \ln\langle e^{G(\Dvec(\xvec), N) - \lambda'^t\cdot\Dvec(\xvec)} \rangle_{\lambda'} 
\nonumber \\
&-& \langle G(\Dvec(\xvec), N) - \lambda'^t\cdot\Dvec(\xvec)\rangle_{\lambda'}
\label{eqn:KL-divergence-RW-from-N}
\end{eqnarray}
where 
\begin{eqnarray}
&&e^{G(\Dvec(\xvec), N)}
\nonumber \\
&=& e^{(\lambda + \Cmat_\lambda^{-1}\cdot(\Dvec_e - \ED_\lambda))^t 
\cdot (I + N\Kmat^{-1}\cdot\Cmat_\lambda^{-1})^{-1}\cdot( \Dvec(\xvec) - \Dvec)}
\nonumber \\
&\cdot& e^{-\frac{1}{2N}\quadratic{(N\Kmat^{-1} + \Cmat_\lambda)^{-1}}{(\Dvec(\xvec) - \Dvec_e)}}
\label{eqn:G-in-KL}
\end{eqnarray}
Eq.~\ref{eqn:KL-divergence-RW-from-N} enables the quantitative
estimation of the difference between the distributions generated by
the restrained ensemble simulations and by the maximum entropy method.
Minimization of $\Delta_{\text{KL}}[p_{\lambda_\text{ME}}||p_N]$ with
respect to $N$ yields the number of replicas to use such that the
restrained ensemble simulations generate a distribution closest to
that of the maximum entropy method.

\section{Results and Discussions}

\begin{figure}
\includegraphics*[width=2.8in]{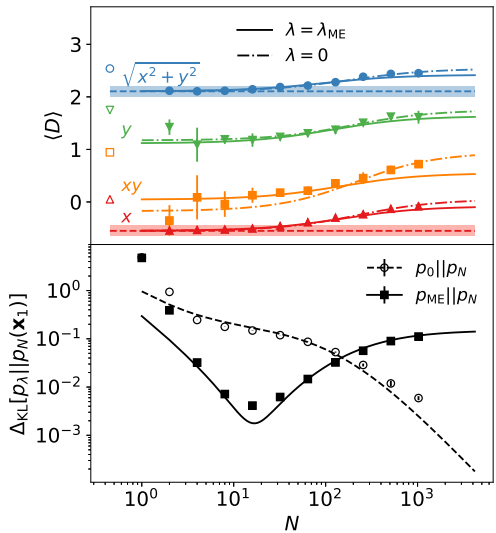}
\caption{The restrained ensemble simulations of a 2-dimensional
  potential at different numbers of replicas. Top: the expected values
  of 4 observables, of which $x$ and $\sqrt{x^2+y^2}$ are restrained
  to the reference values (dashed lines).  The shaded bands indicate
  the hypothetical uncertainties corresponding to $\Kmat$ used in the
  simulations.  Predictions by Eq.~\ref{eqn:D-expected} using $\lambda
  = \lambda_{\text{ME}}$ (solid lines) and $\lambda=0$ (dot-dashed
  lines) are shown.  The predicted $N$ values to use in restrained
  ensemble simulations are insensitive to the $\lambda$ value used: $N
  = 40\pm0$ for $\lambda=\lambda_{\text{ME}}$ and $N = 44\pm 0$ for
  $\lambda = 0 $.  The empty markers indicate the expected values in
  the unbiased simulations.  Bottom: The KL-divergence to the
  distribution of the maximum entropy method (squares) and the
  unbiased distribution (circles) from the marginal distribution of
  the restrained ensemble simulations. The numerical values were
  computed from the simulation samples using a variant of the
  estimators by Wang {\it et al.}~\cite{wang2005divergence}, which
  works for this 2-dimensional system but will become impractical in
  higher dimensions. The lines show the theoretical predictions by
  Eq.~\ref{eqn:KL-divergence-RW-from-N}.
\label{fig:mixed-gaussians}
}
\end{figure}

I demonstrate the validity of Eq.~\ref{eqn:D-expected} and
Eq.~\ref{eqn:KL-divergence-RW-from-N} using the Monte Carlo simulations of
a 2-dimensional potential that generates a distribution of mixed
Gaussians:
\begin{eqnarray}
&&p(x,y; f_a, f_b) = e^{-E(x,y; f_a, f_b)}
\nonumber \\
&=& \sum_{s=a,b}\frac{f_s}{\sqrt{2^2\pi^2|\sigma_s|}}e^{-\frac{1}{2}\quadratic{\sigma_s^{-1}}{(x-x_s, y-y_s)}} 
\label{eqn:mixed-gaussian}
\end{eqnarray}
where $f_a$ and $f_b$ are the mixing coefficients constrained by $f_a
+ f_b = 1$. The parameters used in the simulations are $\sigma_a =
((1.5, 0)^t, (0, 1.5)^t)$, $(x_a, y_a) = (-1, 0)$, and $\sigma_b =
((1.5, 0.1)^t, (0.1, 1.5)^t)$, $(x_b, y_b) = (0.5, 2.5)$. The
restrained ensemble simulations
(Eq.~\ref{eqn:restrained-ensemble-distribution}) and the maximum
entropy method are applied to a potential with $f_a=0.3$, restraining
two observables, $x$ and $\sqrt{x^2 + y^2}$, to reference values that
are equal to the observables' mean values in a reference distribution
that has $f_a = 0.7$.  The restraining strength is $\Kmat =
\text{diag}( 100, 100 )$.

For sufficiently large $N$, Eq.~\ref{eqn:D-expected} predicts the
expected values of both the restrained and unrestrained observables in
quantitative agreement with their mean values in the restrained
ensemble simulations (Fig.~\ref{fig:mixed-gaussians}).  As $N$
increases, the distribution generated by the restrained ensemble
simulations tends to the unbiased distribution
($\lim_{N\rightarrow\infty}\Delta_{\text{KL}}[p_0||p_N] = 0$).
Notably, $\Delta_{\text{KL}}[p_0||p_N]$ is relatively insensitive to
$N$ around the largest value of $N$ for which the expected values of
the restrained observables remain close to their reference values .

My results suggest the following method of setting $N$ so that the
simulations are minimally restrained by experimental data: Iteratively
perform a series of biased simulations of $p_{\lambda_j}(\xvec)$,
updating $\lambda_j$ by either 
\begin{equation}
\lambda_j = \lambda_{j-1} +
\Cmat_{\lambda_{j-1}}^{-1}\cdot( \Dvec_e - \ED_{\lambda_{j-1}} )
\label{eqn:lambda-updates}
\end{equation}
or other update schemes~\cite{stinis2005a} such that
$\lim_{j\rightarrow\infty}\lambda_{j} = \lambda_{\text{ME}}$. For each
iteration, compute $\langle \Dvec(\xvec_1) \rangle_{N,\Kmat,\Dvec_e}$
using Eq.~\ref{eqn:restrained-D-expected} and select the largest $N_j$
so that the expected values of all the restrained observables are
within the measurement uncertainties from their experimental
values. Continue until $N_j$ no longer changes. 

\begin{figure}
\includegraphics*[width=2.7in]{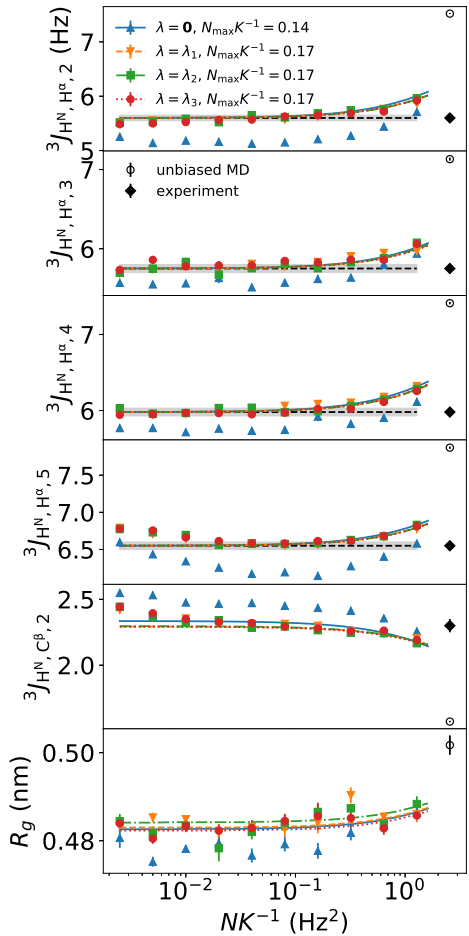}
\caption{ The $^3J$ couplings in and the radius of gyration of
  Ala$_5$.  The experimental values of 4 restrained observables,
  $^3J_{\text{H}^{\text{N}},\text{H}^\alpha,r=2,3,4,5}$, are indicated
  by the dashed lines surrounded by a shaded band, which signifies the
  statistical uncertainty in the measurement, assumed to be
  $\delta=0.05$Hz ($K = 1/\delta^2$).  The lines in each panel
  indicate the expected values of the observables predicted by
  Eq.~\ref{eqn:restrained-D-expected} (for the restrained observables)
  and Eq.~\ref{eqn:unrestrained-D-expected} (for the unrestrained
  observables $^3J_{\text{H}^{\text{N}},\text{C}^\beta,2}$ and $R_g$),
  with different lines corresponding to different values of
  $\lambda_j$, $\langle \Dvec \rangle_{\lambda_j}$ and
  $\Cmat_{\lambda_j}$ used. The discrete symbols indicate the expected
  values calculated from the reweighted simulations
  (Eq.~\ref{eqn:marginal-distribution}) at different values of $N$.
  }
\label{fig:ala5}
\end{figure}

I illustrate this method of setting $N$ by MD simulations of the
polyalanine peptide Ala$_5$ restrained by the $^3J$ couplings
measured by NMR experiments~\cite{Graf:2007,Wickstrom:2009}.  All MD
simulations were performed using the OpenMM
software~\cite{eastman2013}, with Amber99SB-ILDN force
field~\cite{LindorffLarsen:2010} and a generalized Born implcit
solvent model~\cite{onufriev2004exploring}.  Langevin
integrator~\cite{Leimkuhler2015} was used to maintain the temperature
of the molecular system at 300 K.  When Ala$_5$ was simulated without
any restraints, there was substantial discrepancy between the
predictions by the unbiased simulation and the experimental values for the
$^3J$ couplings between H$^\text{N}$ and
H$^\alpha$ (denoted as $^3J_{\text{H}^{\text{N}},\text{H}^\alpha,r}$ for
$r=2,3,4,5$), which have been used to characterize the four backbone
torsions $\phi_{r=2,3,4,5}$ of Ala$_5$~\cite{vogeli2007limits}
(Fig.~\ref{fig:ala5}).  Below, $^3J_{\text{H}^{\text{N}},\text{H}^\alpha,r=2,3,4,5}$ 
were used as the restrained observables.

The same molecular system was then simulated with a series of biasing
potentials (Eq.~\ref{eqn:biased-distribution}), in which $\lambda_j$
in the $j$'th simulation was determined by
Eq.~\ref{eqn:lambda-updates}. Eq.~\ref{eqn:restrained-D-expected} was
used to predict the expected values of the restrained observables
($^3J_{\text{H}^{\text{N}},\text{H}^\alpha,r=2,3,4,5}$) in the
restrained ensemble simulations for different values of $N$, using
the values of $\lambda_j$ and the corresponding $\langle \Dvec
\rangle_{\lambda_j}$ and $\Cmat_{\lambda_j}$ in different biased simulations
as inputs. The largest $N$ ($N_{\text{max}}$) for which the
predicted values of all the restrained observables are within the
statistical uncertainty from the experimental values is determined for 
each $j$. In this case, only 1 iteration of biased simulation was
necessary to have a converged estimate of
$N_{\text{max}}$ (Fig.~\ref{fig:ala5}).

In practice, it is easier to select the number of replicas $N$ based
on the available computing resources (e.g., the number of CPU cores).
Once $N_{\text{max}}$ has been determined by the above procedure, one
can perform the restrained ensemble simulations using $N$ replicas and
the scaled strength of restraints $\Kmat_{\text{min}} =
(N/N_{\text{max}})\Kmat$, because the restrained ensemble simulations with 
the parameters $(N, \Kmat_{\text{min}})$ and with the parameters
 $(N_{\text{max}}, \Kmat)$ generate equivalent marginal distributions 
(Eq.~\ref{eqn:marginal-func-of-NiK}).  

My results suggest that the restrained ensemble simulations at large
$N$ are equivalent to a single-replica simulation of the reweighted
distribution in Eq.~\ref{eqn:marginal-distribution} (which
becomes Eq.~\ref{eqn:marginal-func-of-NiK} as $N\rightarrow\infty$).
In this case, $N$ can be regarded as a parameter that tunes the
strength of the restraints, which are proportional to the
inverse of the covariance of the experimental measurements.  As $N$
tends to infinity, the resulting distribution becomes unbiased.  To
demonstrate this reweighting approach, the Ala$_5$ system was
simulated according to the marginal distribution in Eq. 12 for
different values of $N$, using different values of $\lambda_j$ and the
corresponding $\langle \Dvec \rangle_{\lambda_j}$ and
$\Cmat_{\lambda_j}$ as inputs. For $j\geq 1$, the expected values of
the restrained ($^3J_{\text{H}^{\text{N}},\text{H}^\alpha,r=2,3,4,5}$)
and unrestrained ($^3J_{\text{H}^{\text{N}},\text{C}^\beta,r=2}$ and
the radius of gyration $R_g$) observables in these reweighted
simulations are in good agreement with the corresponding values
predicted by Eq.~\ref{eqn:D-expected} (Fig.~\ref{fig:ala5}).  Such
reweighted simulations may offer a simple alternative to replica-based
restrained ensemble simulations as a way to incorporate into MD
simulations experimental data with their associated measurement
uncertainties.  

\section{Acknowledgments}
H.\,X. thanks Woody Sherman for helpful suggestions on the manuscript.

\section{Notes}
The author declares no competing financial interest.

%

\end{document}